\documentclass[aps,prl,superscriptaddress,twocolumn,floatfix,a4paper]{revtex4}

\usepackage{graphicx,graphics,epsfig}   
\usepackage{dcolumn}   
\usepackage{amsmath} 
\usepackage{verbatim}   
\usepackage{color}     	 
\usepackage{subfigure}  
\usepackage{times,natbib}
\usepackage{amsmath,amsfonts,amssymb,graphics,graphicx,epsfig,color,times,natbib}

\newcommand{\be}{\begin{equation}}
\newcommand{\ee}{\end{equation}}
\newcommand{\ba}{\begin{eqnarray}}
\newcommand{\ea}{\end{eqnarray}}
\newcommand{\ban}{\begin{eqnarray*}}
\newcommand{\ean}{\end{eqnarray*}}

\newcommand{\si}{\sigma}
\newcommand{\braket}[2]{\mbox{$ \langle #1 | #2 \rangle $}}

\newcommand{\ket}[1]{\mbox{$ | #1 \rangle $}}
\newcommand{\bra}[1]{\mbox{$ \langle #1 | $}}

\begin{document}


\title{Measuring small longitudinal phase shifts: weak measurements or standard interferometry?}
\author{Nicolas Brunner}
\affiliation{H.H. Wills Physics Laboratory, University of Bristol, Tyndall Avenue, Bristol, BS8 1TL, United Kingdom}
\author{Christoph Simon}
\affiliation{Institute for Quantum Information Science and Department of Physics and Astronomy,\\
University of Calgary, Calgary T2N 1N4, Alberta, Canada}

\date{\today}


\begin{abstract}
Recently, weak measurements were used to measure small effects that are transverse to the propagation direction of a light beam. Here we address the question whether weak measurements are also useful for measuring small longitudinal phase shifts. We show that standard interferometry greatly outperforms weak measurements in a scenario involving a purely real weak value. However, we also present an interferometric scheme based on a purely imaginary weak value, combined with a frequency-domain analysis, which may have potential to outperform standard interferometry by several orders of magnitude.
\end{abstract}

\maketitle

A cornerstone of quantum mechanics is that a measurement generally perturbs the system. Indeed during the process of a (standard) quantum measurement, the state of the system is projected onto one of the eigenstates of the measured observable. However, in 1988, in the context of foundational research on the arrow-of-time in quantum theory, Aharonov, Albert and Vaidman \cite{WM88} discovered that quantum mechanics offers a much larger variety of measurements. As a matter of fact, the only restriction quantum mechanics imposes on measurements is a trade-off between information-gain and disturbance. Therefore, strong (or standard) quantum measurements are only part of the game. There are also "weak" measurements \cite{weak}, which disturb the system only very little, but which give only limited information about its quantum state.

Weak measurements lead to striking results when post-selection comes into play. In particular, the ``weak value'' found by a weak measurement on a pre- and post-selected system can be arbitrarily large, where the most famous example is the measurement of a spin particle leading to a value of 100 \cite{WM88}. Because of such unorthodox predictions, weak measurements were initially controversial \cite{duck89+leggett89}, and were largely considered as a strange and purely theoretical concept. However, they turn out to be a useful ingredient for exploring the foundations of quantum mechanics. In particular, they bring an interesting new perspective to famous quantum paradoxes, as illustrated by recent experiments \cite{lundeen09} on Hardy's paradox \cite{hardy92,aharonov02}. Furthermore, they also perfectly describe superluminal light propagation in dispersive materials \cite{solli04,superluminal}, polarization effects in optical networks \cite{weakPMD} and cavity QED experiments \cite{wiseman02}. Weak measurements have been demonstrated in numerous experiments \cite{lundeen09,solli04,superluminal,ritchie91}, and were recently shown to be relevant in solid-state physics as well \cite{williams08}.

Already in 1990 Aharonov and Vaidman \cite{WM90} pointed out the potential offered by weak measurements for performing very sensitive measurements. More precisely, when weak measurements are judiciously combined with pre- and post-selection, they lead to an amplification phenomenon, much like a small image is magnified by a microscope. This effect is of great interest from an experimental perspective, since it gives access to an experimental sensitivity beyond the detector's resolution, therefore enabling the observation of very small physical effects. Hosten and Kwiat \cite{hosten,resch} recently used this technique to perform the first observation of the spin Hall effect of light. More recently, Refs \cite{dixon09,starling09} took advantage of the same method to amplify small transverse deflections of an optical beam, in order to measure the angular deflection of a mirror with an impressive resolution of 400 frad.

The effects measured in Refs. \cite{hosten,dixon09,starling09} are all transverse to the light propagation direction. Here we investigate the power offered by weak measurements for measuring small longitudinal phase shifts. We perform a comparison with standard interferometry, which is the natural reference in this context, taking into account the influence of experimental errors.
We consider the situation where the most important errors are not due to statistics (i.e. the total number of photons detected), but to imperfections in the setup. For example, this is true if the phase that is to be measured is stationary (e.g. a weak-contrast, but stable, microscopic sample), making it possible to integrate over arbitrarily long times. While infinitely long integration times are of course an idealization, situations where the precision limit is not set by statistics, but by other factors such as unavoidable alignment errors, are very common in practice.
We first consider a scenario involving a large real weak value (combined with an analysis in the time-domain), and show that it is greatly outperformed by interferometry. Then we present an interferometric setup involving a purely imaginary weak value (combined with an analysis in the frequency-domain), and show that it has potential to outperform standard interferometry by 3 orders of magnitude.

\textit{\textbf{Weak measurements with post-selection.}} We begin with a brief review of weak measurements. We consider a measurement scenario involving a physical system, in a quantum state $\ket{ \psi}$, and a measurement device represented by a pointer state $\ket{g(x)}$, where $x$ is the degree of freedom used by the observer to eventually read out the pointer and $g(x)$ is the associated wave function. The observable to be measured is denoted $A$. For simplicity we choose the system to be a two-level system and write its state in the eigenbasis of $A$: $\ket{\psi}=\alpha \ket{0}+\beta \ket{1}$, where $ |\alpha| ^2 + |\beta|^2=1$, $A\ket{0}=\ket{0}$ and $A\ket{1}=-\ket{1}$.

The first step in the measurement process is an interaction between the system and the pointer, represented by a unitary operation $U = e^{-iH \Delta t}$, with the interaction hamiltonian given by $H = \chi P A$, where $P$ is the momentum operator acting on the pointer state and $\Delta t $ the duration of the interaction. Thus the action of $U$ is to shift the pointer depending on the quantum state $\ket{\psi}$ of the system. More precisely, the initial state $ \ket{g(x)}  \ket{\psi}$ gets mapped to

\ba \ket{\Psi} = U \ket{g(x)} \ket{\psi} =  \alpha \ket{g_+}\ket{0} + \beta \ket{g_-}\ket{1} \ea where $\ket{g_{\pm}} \equiv \ket{g(x\pm \tau)}$ and $ \tau=\chi \Delta t$.

In a second step the observer will perform a read-out of the pointer, by measuring its position, thus obtaining some information about the system. The crucial point is now the ratio between the pointer spread, noted $\si$, and the shift of the pointer $\tau$. On the one hand, if $ \si \ll \tau$, the position of the pointer gives full information about the measurement outcome of $A$, since the overlap $\braket{g_+}{g_-}$ is essentially 0. In this case, the measurement is strong, and corresponds simply to a standard quantum measurement. Note also that in this case the state $\ket{\Psi}$ is maximally entangled (for $|\alpha|=|\beta|$). Thus, measuring the pointer strongly perturbs the state of the system.

On the other hand, when $ \si \gg \tau$, the position of the pointer provides only limited information about the state of the system, since the overlap $\braket{g_+}{g_-}$ is roughly 1. Here the measurement is said to be weak. In this case the state $\ket{\Psi}$ is close to separable for all values of $\alpha$ and $\beta$, thus the state of the system is only weakly perturbed by the pointer measurement.

As mentioned above, weak measurements are of particular interest when combined with post-selection. So we add a post-selection on the state of the system before the observer gets to measure the pointer. The post-selected state is denoted $\ket{\phi}= \mu \ket{0}+ \nu \ket{1}$ with $|\mu|^2+|\nu|^2=1$. Thus, the pointer state is now given by

\ba\label{f(x)} f(x)= \bra{\phi}e^{-i \tau P A } \ket{g(x)} \ket{\psi} = \alpha \bar{\mu} \ket{g_+} + \beta \bar{\nu} \ket{g_-}. \ea In the regime of weak measurements, i.e. $ \si  \gg \tau$, we find 

\ba\nonumber  f(x) &\simeq& \bra{\phi} (\openone -i\tau PA ) \ket{g(x)}\ket{\psi} \\\label{AAV} &=& \braket{\phi}{\psi}
(\openone -i\tau A_w P)\ket{g(x)} \\\nonumber &\simeq& \braket{\phi}{\psi}  e^{-i\tau A_w P} \ket{g(x)}. \ea Here $A_w$ is the so-called weak value \cite{WM88}, given by

\ba\label{Aw} A_w =  \frac{\bra{\phi}A\ket{\psi}}{\braket{\phi}{\psi}},  \ea which should be understood as the mean value of observable $A$ when weakly measured between a pre-selected state $\ket{\psi}$ and a post-selected state $\ket{\phi}$. Note that $A_w$ in general a complex number \cite{Jozsa07}. Eq. \eqref{AAV} shows that the shift in position of the pointer is now amplified by a factor of $\text{Re}[A_w]$. The imaginary part of $A_w$ is associated to a shift of the pointer in momentum space. As seen from Eq. \eqref{Aw}, $A_w$ can be made arbitrarily large, when choosing pre- and post-selected states such that $\braket{\phi}{\psi} \simeq 0$ -- the probability of a successful post-selection becomes then indeed arbitrarily small.  Note that the weak value amplification is obtained via a destructive interference, occurring at the post-selection, between the pointer states $\ket{g_+}$ and $\ket{g_-}$.

\textit{\textbf{Measuring small effects using weak measurements.}} Let us reconsider the above discussion from a different perspective. We now think of the unitary interaction $U$ between the system and the pointer, not as being a measurement process but rather as a small physical effect. For instance, this effect could be a small birefringence in an optical fibre, introducing a weak coupling between the spatial mode of a light pulse and its polarization. Another example is the spin Hall effect of light, where an optical beam is slightly deflected (perpendicularly to its propagation axis) depending on its polarization.

We will focus on the regime where the effect under consideration is so small that the resolution of our detector is not sufficient to distinguish directly between both pointer states $\ket{g_+}$ and $\ket{g_-}$. Remarkably, weak measurements are useful in this situation. Indeed, by carefully choosing the pre- and post-selection, it is possible to amplify the shift of the pointer; the amplification factor being given by the weak value $A_w$. The price to pay is that the signal intensity will be lowered, since the weak value becomes large in a regime where the pre- and post-selection are almost orthogonal. Still, the possibility to increase resolution while decreasing the signal intensity has already been proven useful for demonstrating tiny transversal effects, such as the Hall effect of light \cite{hosten}, and remarkably small beam-deflections \cite{dixon09,starling09}.

\textit{\textbf{Measuring small longitudinal phase shifts.}} We now investigate the power offered by weak measurements for measuring small longitudinal effects, by performing a comparison with standard interferometry. For weak measurements, we consider two different scenarios, involving first a purely real weak value, and second a purely imaginary weak value. For clarity, we focus on the example of a birefringent element, but our analysis also applies to longitudinal phase shifts that depend on other degrees of freedom (e.g. the path in an interferometer). Specifically our goal is to perform an absolute measurement of a small phase shift induced by the birefringence. As discussed above, we consider the situation where the main experimental limitation comes from errors in the alignment of the setup as well as from imperfections of the optical elements, rather than from statistics (i.e. photon shot noise) \cite{putman,higgins}.

\begin{figure}[b]
  \includegraphics[width=0.96\columnwidth]{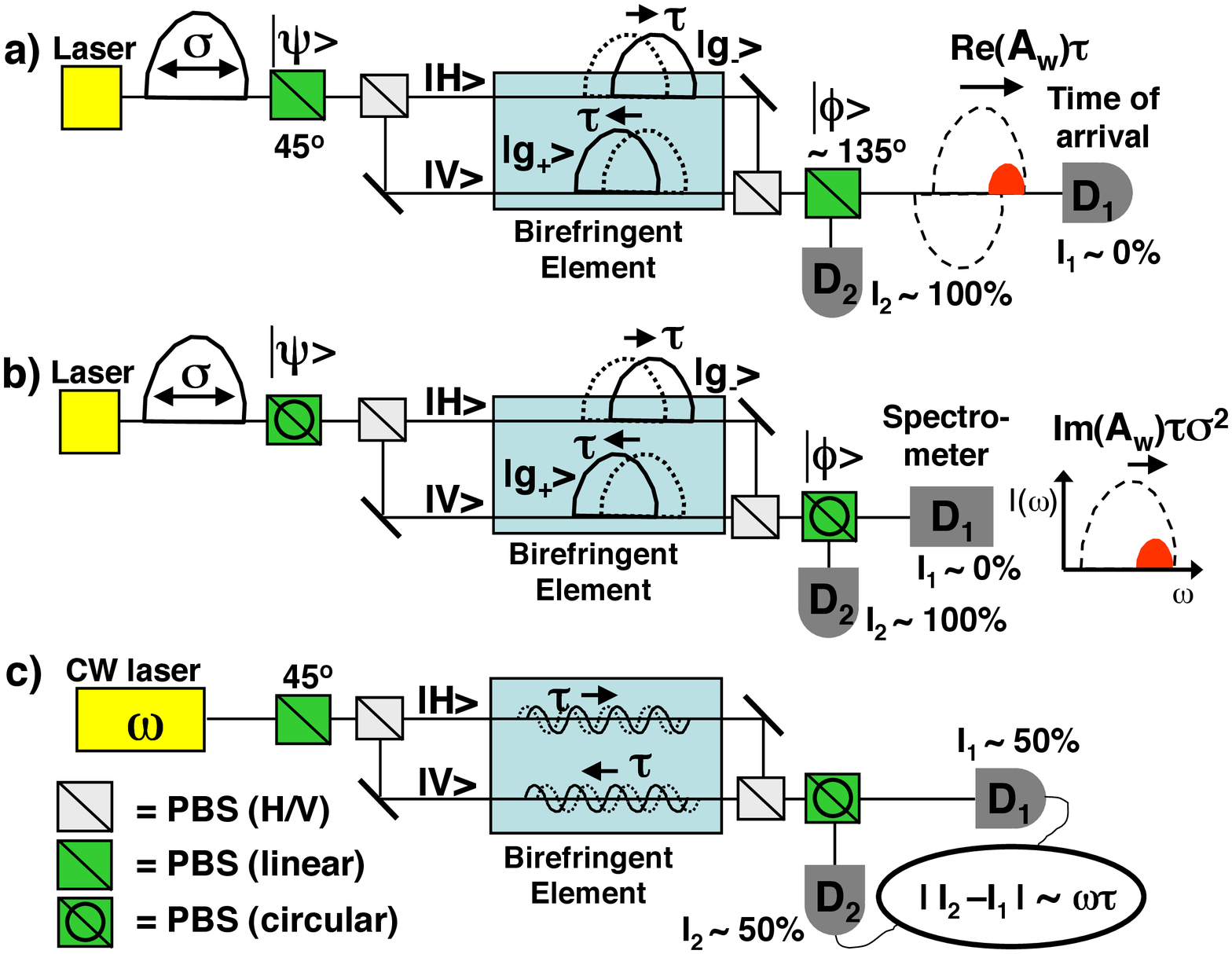}
  \caption{Measuring a small longitudinal effect (here a tiny birefringence) using: a) weak measurements with a real weak value, b) weak measurements with an imaginary weak vale, c) interferometry. Each technique is characterized by a particular alignment of the interferometer and by a specific type of measurement. For weak measurements, the interferometer is set such that the output mode $D_1$ is a dark port -- by choosing (almost) orthogonal pre-selection $\ket{\psi}$ and post-selection $\ket{\phi}$. This achieves an amplification of the small delay $\tau$ introduced by the birefringent element between the horizontally and vertically polarized components. In the case of a real weak value, this amplification occurs in the time-domain; thus we perform a time-of-arrival measurement which gives access to $\tau$. In the case of an imaginary weak value, the pointer is shifted in frequency; the final measurement is thus carried out in the frequency-domain. For interferometry we use a CW laser. The interferometer is aligned such that, when $\tau=0$, the output intensities in modes $D_1$ and $D_2$ are equal. Then, the intensity difference between modes $D_1$ and $D_2$ gives access to the small phase shift $\tau$.}
\end{figure}

\textit{a. Weak measurements, real weak value.} We consider the setup of Fig.~1a. The goal is to measure a small longitudinal delay (small phase shift) $\tau$ introduced by the birefringent element between the two pointer states $\ket{g_+}$ and $\ket{g_-}$ which correspond to the element's (orthogonal) polarization eigenmodes; here we choose $\ket{0}=\ket{H}$ and $\ket{1}=\ket{V}$. The delay $\tau$ slightly modifies the time of arrival of the pulse, depending on its polarization state. Using polarizing beam splitters (PBS), we perform a pre- and a post-selection, which result in an amplification of the delay -- the amplification factor being $\text{Re}[A_w]$. Typically we choose $\ket{\psi}=\frac{1}{\sqrt{2}}(|0\rangle+|1\rangle)$ and $\ket{\phi}\approx \frac{1}{\sqrt{2}}(|0\rangle-|1\rangle)$, resulting in a large (purely) real weak value. Finally, in order to get access to the delay $\tau$, we measure the time-of-arrival of the pulse in mode $D_1$, which is here a dark port since $\braket{\phi}{\psi}\approx 0$.

Indeed the probability to obtain a successful post-selection is given by $p= n/N=|\braket{\phi}{\psi}|^2$, where $N$ is the number of photons initially prepared in the state $\ket{\psi}$, and $n$ is the number of photons actually detected (i.e. passing the post-selection). Thus we have that $A_w \simeq \frac{1}{\sqrt{p}}$. Note that since there are no non-linear effects, it does not matter whether the photons are sent through the setup one by one or not. The shift of the pointer is then given by $\tau A_w \simeq \tau / \sqrt{p}$. Clearly, in order to detect the small effect $\tau$, the pointer shift must be larger than the temporal resolution $\Delta t$ of the detector, i.e. $\tau A_w > \Delta t$, implying that $ \tau > \Delta t \sqrt{p}$. Thus, the resolution of the detector is effectively increased by a factor $1/\sqrt{p}$, when photons pass the post-selection with probability $p$.

However, in practice this factor cannot be made arbitrarily large because of experimental imperfections, which lead to the detection of photons in the wrong output port. To be specific, we will focus on errors due to the misalignment of the PBS's; we note $\epsilon$ the error on the angle between both PBS's. Here this forces us to work in the regime where $p>\epsilon^2$, since we monitor the dark port of the interferometer (corresponding to a minimum of intensity). Thus the resolution limit for weak measurements with a real weak value is given by

\ba\label{W_noise} \tau > \epsilon \Delta t. \ea

\textit{b. Weak measurements, imaginary weak value.} We consider the same setup as above. To switch from a purely real to a purely imaginary weak value, it suffices to modify the pre- and post-selection, i.e. the alignment of both PBS's. Here we choose $\ket{\psi}=( \ket{H}+i \ket{V})/\sqrt{2}$ and $\ket{\phi}=(ie^{i\varphi} \ket{H}+e^{-i\varphi} \ket{V})/\sqrt{2}$. The probability to pass the post-selection is then $p=\sin^2{\varphi}$ and the weak value is $A_w=i \cot{\varphi}$. Finally we perform a measurement of the frequency spectrum of the pointer which gives access to the phase. Here the weak value formalism predicts a displacement of the pointer in the frequency-space $\delta \omega= 2\tau \text{Im}[A_w]/\sigma^2 \simeq 2 \tau / (\varphi \sigma^2)$ for small $\varphi$ \cite{Jozsa07}, which we derive explicitely below.

For simplicity we consider a gaussian pointer $g(t)= C e^{-(t/2\sigma)^2}$ where the constant $C$ ensures that $G(t)=|g(t)|^2$ is a probability distribution. After the post-selection, the pointer is given by Eq. \eqref{f(x)} with $\alpha \bar{\mu}=-ie^{-i\varphi}/2$ and $\beta \bar{\nu}=ie^{i\varphi}/2$. The spectrum of the pointer is
\ba\label{spectrum} F(\omega)= |f(\omega)|^2 = \sin^2(\omega\tau - \varphi)|g(\omega)|^2,\ea where $g(\omega)$ is the Fourier transform of $g(t)$. Based on Eq. \eqref{spectrum}, one can show that for small $\varphi$ the frequency shift is given by

\ba \delta \omega = \frac{\int_{-\infty}^{\infty} \omega F(\omega)d\omega}{\int_{-\infty}^{\infty} F(\omega)d\omega} \simeq  \frac{2 \tau}{ \sigma^2 \varphi }. \ea

In order for this frequency shift to be measurable, we require that $\delta \omega > \Delta \omega $, where $\Delta \omega$ is the resolution of the spectrometer. Considering again an error level of $\epsilon$, we require that $p\simeq \varphi^2 >\epsilon^2$ (as for case a)). Thus the resolution limit for weak measurements with an imaginary weak value is given by

\ba\label{WI_noise} \tau >  \epsilon \sigma^2 \Delta \omega /2 . \ea

\textit{c. Interferometry.} Here we consider a standard interferometric scheme (see Fig.~1b). It is now advantageous to use a laser of frequency $\omega$ working in continuous wave (CW) mode. The light is first polarized at $45^o$. The final PBS operates in the circular polarization basis, corresponding to the states $\frac{1}{\sqrt{2}}(|0\rangle \pm i|1\rangle)$, such that without the delay (i.e. for $\tau=0$), an equal intensity is obtained in detectors $D_1$ and $D_2$. Thus, for small $\tau$, one gets $I_{1} \simeq \frac{N}{2}(1+2\omega \tau)$ and $I_{2} \simeq \frac{N}{2}(1-2\omega \tau)$. The intensity difference gives then access to the phase:
\ba |I_{2}-I_{1}|= 2 N \omega \tau. \ea Again assuming an error $\epsilon$ on the angle between both PBS's, giving here of order $N \epsilon$ potentially erroneous detections in each detector, we are led to require that $\omega \tau > \epsilon$; note that the error level depends now linearly on $\epsilon$ since the interferometer operates in the regime where $I_1\approx I_2$. Thus the resolution limit for interferometry is given by
\ba\label{IF_noise} \tau > \frac{\epsilon}{\omega}. \ea

\textit{\textbf{Comparison.}} From inspection of Eqs. \eqref{W_noise}, \eqref{WI_noise} and \eqref{IF_noise} we see that all techniques discussed above perform equally well from the point of view of robustness to noise; in each case the resolution limit scales linearly with the alignment error $\epsilon$. It is thus essential to compare the numerical factors.

Comparing Eqs. (\ref{W_noise}) and (\ref{IF_noise}) clearly shows that interferometry outperforms weak measurement with a real weak value, since $\Delta t$ is in practice many orders of magnitude greater than $1/\omega$. Whereas the latter is less than 1 fs for optical frequencies, the best currently available single-photon detectors still have $\Delta t>10$ ps \cite{goltsman}, the two timescales are thus separated by more than four orders of magnitude.

However, the weak measurement scenario becomes much more interesting for the case of an imaginary weak value combined with frequency-domain analysis. Currently available lasers can generate femtosecond pulses; e.g. for a Titanium-Sapphire laser (operating at the wavelength $\lambda=700\text{ nm}$) one can have $\sigma =5\text{fs}$ \cite{5fs}. Currently available spectrometers have a spectral resolution of $\Delta \lambda = 5 \text{ pm}$, which leads to a frequency resolution of $\Delta \omega = 20 \text{ GHz}$ at $\lambda=700\text{ nm}$. Overall this leads us to expect an improvement of three orders of magnitude compared to standard interferometry: here we have that $\sigma^2 \Delta \omega = 0.5\text{ as}$, whereas for interferometry we have $1/\omega = 0.4$ fs at $\lambda=700$ nm.

\textit{\textbf{Conclusion.}} We considered the task of measuring small longitudinal phase shifts. Specifically, we studied two techniques based on weak measurements and compared them with standard interferometry, the natural reference in this context. We found that in the case of a real weak value, weak measurements cannot compete with interferometry. However, we also proposed a scheme involving an imaginary weak value, combined with frequency-domain analysis, which could in principle outperform interferometry by three orders of magnitude. Let us note that, while our technique is inspired by the quantum formalism of weak values, it also works for classical light.

Here we focused on the case where the resolution is limited by alignment errors (which can account for a finite extinction ratio of the polarizers, for example), but it would also be interesting to study the influence of other types of experimental errors (e.g. detector dark counts). Another important open question is whether weak measurements could also enhance the resolution in situations where statistics is the limiting factor.

Finally, we point out that there exist situations involving longitudinal effects in which weak measurements with real weak values can nevertheless be useful. This is the case in differential interference (DIC) microscopy \cite{DIC} which turns out to be strongly connected to weak measurements (see Appendix).

\emph{Acknowledgements.} We thank N. Gisin and S. Popescu for discussions. N.B. acknowledges financial support from the UK EPSRC.

\section{Appendix: DIC microscopy and weak measurements}


Finally we point out a connection between differential contrast (DIC) microscopy \cite{DIC} and weak measurements, which shows that weak measurements with a real weak value can nevertheless be useful in situations involving small longitudinal effects. The goal of DIC microscopy is not to obtain the highest possible resolution, but rather a good {\it visual} contrast. From Fig.~2 it is clear that we recover in DIC microscopy the three fundamental steps of the weak measurement scenario: 1) pre-selection, 2) weak measurement, 3) post-selection. Notably, the pre- and post-selection correspond to crossed polarizers, thus resulting in a dark port. By monitoring this dark port, DIC microscopy provides the observer with an enhanced contrast in transparent samples; it gives three-dimensional physical relief corresponding to the minute variations of optical density in the sample. Note that using a dark port is advantageous here because the human eye is a threshold detector \cite{rieke}, for which it is much easier to detect small fluctuations compared to a zero background intensity, rather than the difference of two strong signals as in standard interferometry. On the other hand, our preceding analysis indicates that when photon counting or linear intensity detectors are used, a standard interferometric setup will provide much better resolution than DIC microscopy.

\begin{figure}[b]
  \includegraphics[width=\columnwidth]{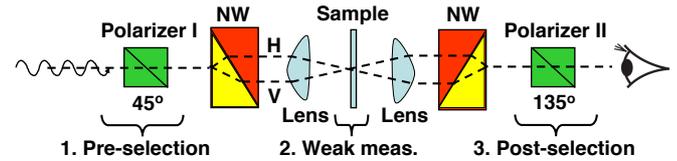}
  \caption{Setup of DIC microscopy and connection to weak measurements. A light beam is polarized at $45^o$ by polarizer I; this is the pre-selection. Next, a Nomarski-Wollaston (NW) prism (basically a PBS) separates horizontal and vertical polarization modes in two parallel beams. The beams travel through different areas of the sample, and thus acquire slightly different phases due to local differences in the refractive index or thickness of the sample. These tiny anisotropies of the sample perform the weak measurement. After the sample, a second NW prism recombines both polarization beams. Finally, polarizer II post-selects the $135^o$ polarized component, which will be monitored; this is the post-selection. Note that the pre- and post-selection correspond to crossed polarizers, as in the weak measurement scenario.}
\end{figure}

\end{document}